\documentclass{article}
\usepackage[T1]{fontenc}
\usepackage[utf8]{inputenc}
 \usepackage{ismir}
\usepackage{amsmath,amssymb,cite,url}
\usepackage{xcolor}
\usepackage{graphicx}
\usepackage{booktabs}
\usepackage{listings}
\usepackage{tikz}
\usepackage{pifont}

\lstdefinestyle{faust}{
  basicstyle=\ttfamily\small,
  commentstyle=\color{gray},
  keywordstyle=\bfseries,
  numbers=none,
  frame=single,
  xleftmargin=1em,
  xrightmargin=1em,
}

\title{PEACE: Joint Embeddings of DSP Effects Code and Audio}

\oneauthor
  {David Braun and Adam Finkelstein}
  {Princeton University\\\texttt{db1224@princeton.edu}}

\def\authorname{D. Braun and A. Finkelstein}

\usepackage[bookmarks=false,pdfauthor={\authorname},pdfsubject={\pdfsubject},hidelinks]{hyperref}

\begin{document}

\maketitle

\begin{abstract}
This paper introduces PEACE, the first joint embedding of audio effect code and output audio.
Building on SLAP's multimodal objective, we pair an AFx-Rep audio encoder with two code encoders for Faust, a functional language for audio signal processing.
First, we evaluate a fine-tuned T5 transformer over Faust source code.
Second, we evaluate a message-passing graph neural network over an intermediate representation of the Faust compiler, capturing both topology and UI parameters.
We evaluate on audio-to-code retrieval, where masking UI parameters at inference yields embeddings that encode effect chain topology alone.
When parameters are visible, the two code encoders tie on retrieval of mixed-length chains but have tradeoffs on single-effect galleries.
With parameters fully masked, PEACE recovers ordered chain topology far above chance without the limitations of supervised methods.
PEACE outperforms pretrained models on an out-of-distribution reverb retrieval benchmark and can improve frozen audio-only representations.
Its dual understanding of topology and parameters lays the groundwork for music information retrieval systems that search, generate, and condition on DSP code.

\end{abstract}

\section{Introduction}
Software audio effects such as C++ plugins are everywhere in modern music production~\cite{steinmetz_audio_2025}, but no multimodal embedding connects effected audio back to the code that produced it.
Joint embeddings have closed analogous cross-modal gaps: CLIP~\cite{radford_learning_2021} for images and text, CLAP~\cite{elizalde_natural_2024, wu_large-scale_2023} for audio and text.
Each has spawned countless applications across retrieval, classification, and generation.
What would a joint embedding of audio and DSP code enable?

Prior work on audio effects covers pieces of this problem.
Text-guided methods~\cite{chu_text2fx_2025, doh_can_2025, doh_llm2fx-tools_2026} rely on natural language, but even specific descriptions like ``shimmery reverb in a cold cavern'' might oversimplify an effect, losing details that distinguish configurations.
Human labels are also costly and limit available data.
Direct regression of effect parameters is ill-posed, since many distinct settings can produce indistinguishable output, such as reordering time-invariant effects~\cite{smith_introduction_2007};
for relevance to synthesizers, see~\cite{hayes_audio_2025}.
Classification~\cite{guo_automatic_2023, wada_hyperbolic_2025, rossi_automatic_2025, okita_audio_2026} and generative~\cite{lee_blind_2023, lee_reverse_2025} approaches avoid ill-posedness but produce graphs or a fixed set of labels rather than embeddings where programs can be compared.
Audio encoders~\cite{steinmetz_st-ito_2024, yeh_fx-encoder_2025} recover effect-based representations from wet signals but don't enable a similarity metric for programs or code-to-audio retrieval.

To build such an embedding, the code modality must capture both program structure and parameter values.
Faust~\cite{orlarey_faust_2009} is a domain-specific language for audio signal processing whose Block Diagram Algebra~\cite{orlarey_syntactical_2004} gives programs a graph structure suitable for encoding.
Faust has extensive libraries and compiles to C, C++, LLVM, JAX, Wasm, and more, giving it broad deployment reach.

We propose PEACE (\textbf{P}airwise \textbf{E}ffected \textbf{A}udio \textbf{C}ode \textbf{E}mbeddings), which learns joint embeddings of Faust effect code and audio using the pairwise objective from Siamese Language-Audio Pretraining (SLAP)~\cite{guinot_slap_2025}.
The paper makes the following contributions:
\begin{itemize}
    \item With SLAP, we pair an AFx-Rep audio encoder~\cite{steinmetz_st-ito_2024} with two complementary code encoders for Faust: a fine-tuned T5 transformer~\cite{raffel_exploring_2023, roberts_scaling_2023} over Faust source, and BoxGraph, a smaller graph neural network (GNN) over the Block Diagram Algebra.
    \item T5 and BoxGraph roughly tie on cross-modal retrieval from galleries spanning chain lengths 1--3, and T5 generalizes better to even longer chains. On single-effect galleries, they split wins across effect types.
    \item PEACE (especially with BoxGraph) outperforms pretrained AFx-Rep on a reverb retrieval benchmark where audio comes from professional plugins not seen during training. Cross-modal SLAP training also improves the performance from a frozen pretrained AFx-Rep.
    \item Finally, by masking effect parameters in code embeddings, PEACE lets audio queries retrieve chain topologies, with BoxGraph outperforming T5.
\end{itemize}

We hope PEACE lays the groundwork for improvements in style transfer, dry stem recovery, and DSP code generation.
We release our codebase, model weights, and an interactive tool for exploring multimodal embeddings.\footnote{\url{https://github.com/DBraun/PEACE}}

\section{Related Work}

\textbf{Audio Effect Representations.}
Audio foundation models lack nuanced understanding of audio effects because effects are commonly used as training augmentations, making models invariant to them~\cite{steinmetz_audio_2025}.
ST-ITO~\cite{steinmetz_st-ito_2024} trains an effect-aware encoder, AFx-Rep, on a supervised task classifying which individual effect and parameter configuration were applied to audio.
Fx-Encoder++~\cite{yeh_fx-encoder_2025} adds the ability to extract instrument-specific effect representations from full mixtures according to CLAP-space queries.

\textbf{Effect Chain Estimation.}
Prior methods estimate effect chains via supervised classification~\cite{guo_automatic_2023, wada_hyperbolic_2025, rossi_automatic_2025, okita_audio_2026}, autoregressive graph generation~\cite{lee_blind_2023}, graph-based pruning~\cite{lee_reverse_2025}, and iterative inversion with dry-signal estimation~\cite{take_audio_2024}.
Okita and Katayose~\cite{okita_audio_2026} combine several of these ideas.
A DNN predicts the set of effect types and the dry signal.
Then, the same evolutionary strategy of ST-ITO searches over effect orderings and parameters by rendering each candidate and scoring wet signal reconstruction.

\textbf{Text-Guided Audio Effects.}
Text-based approaches use natural language to control effect parameters, either via optimization in CLAP-space~\cite{chu_text2fx_2025, ki_fxsearcher_2025} or LLM-based parameter prediction~\cite{doh_can_2025, doh_llm2fx-tools_2026}.
However, joint language-audio embeddings weakly encode perceptual timbre.
Even the top performer, LAION-CLAP, inverts the human-labeled direction of ``warm'' for both EQ and reverb~\cite{deng_joint_2025}.

\textbf{Representation Learning for Code.}
GraphCodeBERT~\cite{guo_graphcodebert_2021} augments BERT-style pretraining with data flow graphs and graph-guided masked attention to capture variable-level semantics, while CodeT5+~\cite{wang_codet5_2023} is an encoder-decoder code model pretrained on multiple programming languages.
Kouteili et al.~\cite{kouteili_embedding_2025} learn joint embeddings of Sonic Pi code and audio, focusing on programs which generate sound rather than process it.
Their text encoder (DistilRoBERTa) yields cosine similarity above 0.995 even when parameters change substantially, because tokenizers tend to fragment numbers awkwardly (e.g., \texttt{"0.3141"} $\to$ \texttt{["0.3","141"]}).
Prior related work for Faust covers code generation~\cite{ly_digital_2023, lawrence_synthesis_2023} and genetic-algorithm effect matching~\cite{chowdhury_crossroads_2019, kouteili_crossroad2_2025} but not representation learning.

\section{Method}

Given pairs $(c, x_{\text{wet}})$ of Faust code and stereo wet audio, we learn encoders $\mathcal{E}_C$ and $\mathcal{E}_A$ that map code and audio to a shared $d$-dimensional space.
At inference, an audio (or code) query is compared against a preexisting gallery of code (or audio) embeddings by cosine similarity.

\subsection{Training Objective: SLAP}

\begin{figure}[t]
\centering
\includegraphics[width=\columnwidth]{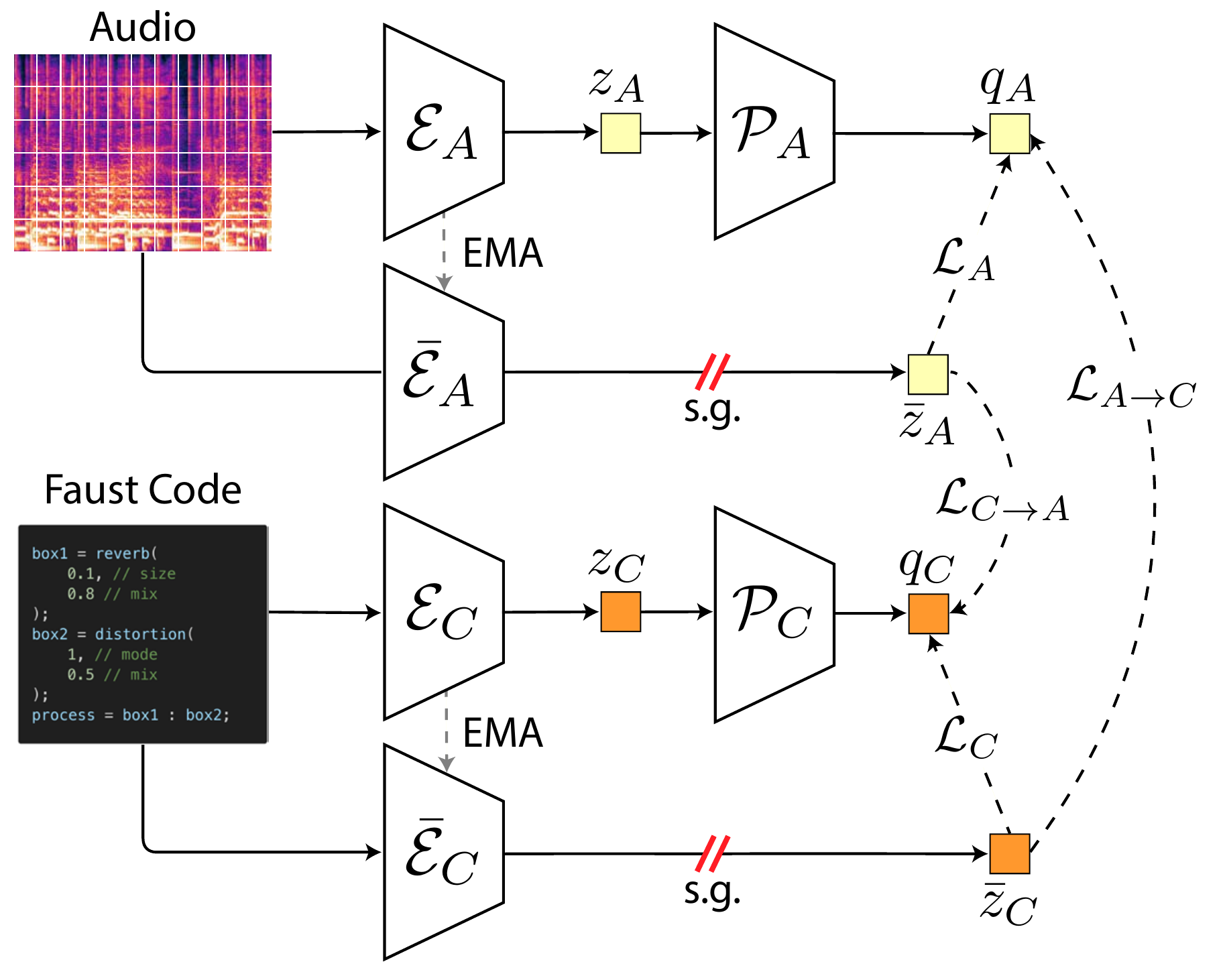}
\caption{
SLAP architecture.
Each modality's online branch pairs encoder-projector $\mathcal{E}_M$ (producing $y_M$ then $z_M$) with predictor $\mathcal{P}_M$ to produce $q_M$.
    The EMA target branch $\bar{\mathcal{E}}_M$ produces stop-gradient $\bar{z}_M$ with no predictor.
The $\mathcal{L}$ terms show loss gradients flowing through the predictors into the online branches, not the EMA targets.
}
\label{fig:slap_architecture}
\end{figure}

Our objective follows SLAP~\cite{guinot_slap_2025} (\figref{fig:slap_architecture}).
Unlike contrastive approaches such as CLAP~\cite{elizalde_natural_2024, wu_large-scale_2023}, SLAP's non-contrastive BYOL objective~\cite{grill_bootstrap_2020} doesn't need negative-samples in batches and allows for gradient accumulation.
SLAP also achieves a smaller modality gap~\cite{liang_mind_2022, fahim_its_2024}, which in prior work benefited text-to-audio generation~\cite{nistal_improving_2024}.

In our setting, for each modality $M \in \{A, C\}$, an encoder-projector $\mathcal{E}_M$ produces a modality-specific latent $y_M$ followed by a projection $z_M \in \mathbb{R}^{d}$.
The \textit{online} branch additionally passes $z_M$ through a predictor $\mathcal{P}_M$ to produce $q_M \in \mathbb{R}^{d}$.
The \textit{EMA target} branch $\bar{\mathcal{E}}_M$ produces only a stop-gradient projection $\bar{z}_M$, with no predictor.
All $z$ and $q$ are $L_2$-normalized.
Let $\theta_M$ denote the parameters of $\mathcal{E}_M$ and $\bar{\theta}_M$ those of $\bar{\mathcal{E}}_M$, which track them via exponential moving average with $\tau = 0.98$:
\begin{equation}
    \bar{\theta}_M \leftarrow \tau \bar{\theta}_M + (1-\tau) \theta_M.
\end{equation}
The loss sums four cosine distances between online predictions $q_M$ and target projections $\bar{z}_{M'}$, with two intermodal terms ($M \neq M'$) and two intramodal terms ($M = M'$):
\begin{equation}
    \mathcal{L} = \lambda \big(\mathcal{L}_{A \to C} + \mathcal{L}_{C \to A}\big) + (1-\lambda) \big(\mathcal{L}_{A} + \mathcal{L}_{C}\big)
\end{equation}
We set $\lambda = 0.5$, which was shown to be optimal in~\cite{guinot_slap_2025}.

\subsection{BoxGraph}
\label{sec:code_encoder}

\begin{figure}[t]
\centering
\includegraphics[width=1.0\columnwidth]{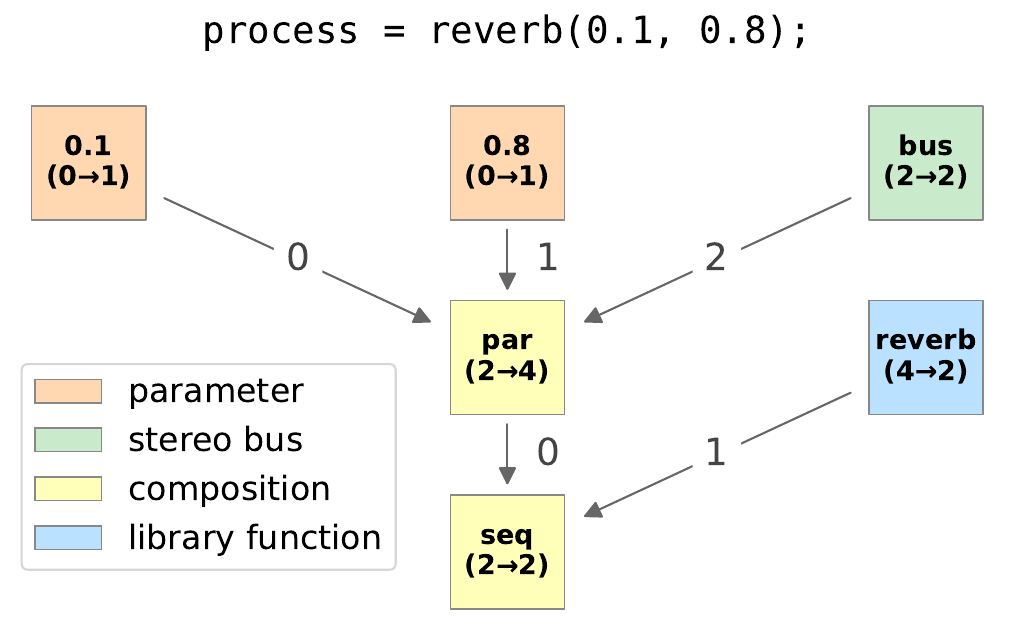}
\caption{
BoxGraph of a stereo effect where \texttt{reverb} is defined in the library.
Each node shows its I/O signature ($X{\to}Y$ means $X$ inputs, $Y$ outputs), edge labels indicate argument positions, and a composition node's signature is fully determined by the operator and children's signatures.
The graph is not a traditional ``audio processor graph'' but instead a composition of functional transformations, each resulting in a new function-like box.
}
\label{fig:boxgraph_example}
\end{figure}

We encode Faust code with a GNN operating on a \textit{BoxGraph}, whose nodes are boxes from the Faust compiler's Box API.\footnote{\url{https://faustdoc.grame.fr/tutorials/box-api/}}
A box is a signal-processing \textit{function} with known input and output channel counts.
A constant $0.1$ is $0{\to}1$, the wire \texttt{\_} is $1{\to}1$, a stereo bus \texttt{\_,\_} is $2{\to}2$, and a reverb taking 2 parameters and a stereo bus is $4{\to}2$.
Boxes compose via Faust's Block Diagram Algebra (BDA)~\cite{orlarey_syntactical_2004}: \texttt{par} puts the reverb's two parameters and the stereo bus in parallel ($2{\to}4$), then \texttt{seq} sequences the result with the reverb ($4{\to}2$), yielding the full $2{\to}2$ effect (\figref{fig:boxgraph_example}).

BDA has five composition operators.
We use only \texttt{par} and \texttt{seq}, leaving \texttt{split}, \texttt{merge}, and \texttt{recurse} to future work.
Note that \texttt{recurse} would not induce cyclicality in the graph because the operators transform function-like boxes.
A later stage of the Faust compiler resolves the recursions with one-sample delays, resulting in code more akin to an audio processor graph.

\textbf{Graph structure.}
The BoxGraph is a directed acyclic graph $G = (V, E)$ where each node $v \in V$ represents a box.
\textit{Composition nodes} combine children using BDA operators.
Edges point from children to parents ($\mathcal{N}(v)$ denotes $v$'s children).
The root is the complete program.
\textit{Terminal nodes} are leaves: parameters, wires, buses, and \textit{DSP nodes} (user-defined functions like \texttt{reverb} or \texttt{delay}), which are box-diagrams but which we treat as atomic primitives.
Any non-terminal subgraph can be promoted to a terminal node by defining it as a library function.

\textbf{Node features.}
All embeddings have dimension $d_{\text{graph}} = 1024$.
Each node's initial feature $x_v$ combines up to three embeddings: a type embedding $E_{\text{type}}[t_v]$ (all nodes);
a DSP vocabulary embedding $E_{\text{dsp}}[d_v]$ (indexing \texttt{reverb} etc.);
and a value embedding $E_{\text{value}}$ (parameter nodes), projecting floats $p_v \in [0,1]$ linearly after remapping to $[-1,1]$, or indexing a per-parameter embedding table for discrete values.
Each discrete parameter slot in the effect library (e.g., \texttt{filter.mode} with 18 modes) has its own $\mathbb{R}^{(V_k+1) \times d_{\text{graph}}}$ table with one extra row reserved for a \texttt{[MASK]} token.
Parameter nodes project $[E_{\text{type}};
E_{\text{value}}]$, DSP nodes project $[E_{\text{type}};
E_{\text{dsp}}]$, and other terminals use only the type embedding.

\textbf{Parameter masking.}
During training, we mask each parameter value with probability 10\%, replacing it with a shared \texttt{[MASK]} vector (floats) or the \texttt{[MASK]} row of its own table (integers).
At inference time, we can intentionally mask all parameters, seeking an embedding aware of only the selection of effects and their ordering.

\textbf{Message passing.}
Initial features $x_v \in \mathbb{R}^{d_{\text{graph}}}$ are LayerNorm'd.
Then $L=4$ message-passing layers with shared weights and pre-LayerNorm are applied.
Edges into composition nodes carry a learned position embedding $e_{\text{pos}}(i) = E_{\text{pos}}[i]$ for the child's argument index $i$.
Writing $c(u)$ for the category of node $u$ (parameter, DSP, or composition) and $h_u^{(\ell)}$ for its features at layer $\ell$, with $h_u^{(0)} = \mathrm{LN}(x_u)$, each layer \textit{messages}, \textit{aggregates}, and \textit{updates}:
\begin{align}
a_v &= \max_{u \in \mathcal{N}(v)}\ W_{\text{cat}}\big[\,W_{c(u)} h_u^{(\ell)} \,;\, e_{\text{pos}}(i_{u \to v})\,\big] \\
h_v^{(\ell+1)} &= \mathrm{FFN}\Big(\mathrm{ReLU}\big(W_{\text{up}}[\,\mathrm{LN}(x_v) \,;\, a_v\,]\big)\Big) \label{eq:mp_update}
\end{align}
Here $\max$ is elementwise, $[\,\cdot\,;\cdot\,]$ is concatenation, and the FFN is a one-hidden-layer MLP with a residual connection.
Unlike standard GNN updates, \eqref{eq:mp_update} \textit{replaces} the hidden state each layer while re-injecting $x_v$, so the update depends only on $x_v$ and the current messages, not on the previous hidden state.
To encode graphs larger than those seen during training, we grow $L$ to the input graph's diameter.
With $L$ at least the diameter, information reaches every node, and extra layers don't change the outputs.

\textbf{Pooling.}
Let $h_v^{(L)}$ denote node $v$'s feature after $L$ layers and let $V_s \subset V$ be the subset consisting of \texttt{seq} composition nodes and DSP nodes;
\texttt{par} nodes and non-DSP terminal nodes are excluded.
The graph embedding is the mean over $V_s$, projected to $y_C \in \mathbb{R}^{512}$:
\begin{equation}
    y_C = \text{Linear}\Big(\frac{1}{|V_s|} \sum_{v \in V_s} h_v^{(L)}\Big).
\end{equation}
Restricting to $V_s$ outperformed the mean or sum of all nodes, as well as taking the root node's final features.

\section{Experiments}

\subsection{Dataset}
\label{sec:dataset}
We build 200K training, 8{,}192 validation, and 4{,}096 test audio-code pairs.
For training and validation, dry audio spans nine categories (bass, drums, guitar, instruments, multitrack stems, music, piano, singing, speech) from 23 public datasets~\cite{riley_filobass_2023, stein_2023_7544032, salimi_2020_3994999, gillet_2006_7432188, hegel_pedroza_2022_7044411, pedroza_2024_11406378, xi2018guitarset, engel_neural_2017, MUSDB18HQ, SiSEC16, pereira2023moisesdb, bogdanov2019mtg, bozkurt_2018_4301737, abesser_2023_7544248, muller_2024_13753319, rong_gong_2022_6536490, zhang2022msinger, soonbeom_choi_2021_4916302, duan_nus_2013, takamichi_jsut, yamagishi2019cstr, richter2024ears, aihub2022multisinger}.
Test data follows AFx-Rep~\cite{steinmetz_st-ito_2024} with drums (IDMT-SMT-Drums~\cite{weber_2023_7544164}), guitar (IDMT-SMT-Guitar~\cite{kehling_2023_7544110}), singing (VocalSet~\cite{wilkins_vocalset_2018}), and speech (DAPS~\cite{mysore_can_2015}).

We use 21 Faust effect implementations with 1--18 parameters each.
Continuous parameters are normalized to $[0,1]$ and discrete parameters are whole numbers.
Per effect, we randomly sample parameters until 30K/4{,}096/4{,}096 presets (train/val/test) each induce a loudness change within $\pm 35$ LU delta (ITU-R BS.1770-4~\cite{series2011algorithms}) on a white-noise reference whose LUFS is sampled from $\mathcal{U}(-30, -10)$.
We render 10-second 48\,kHz clips through 1--3 sequential effects (length uniform) with DawDreamer~\cite{braun_dawdreamer_2021}, rejecting any with LUFS outside $[-40, 40]$ until we accumulate 200K/8{,}192/4{,}096 pairs.
Wet audio is normalized to $-18$ LUFS so the encoder cannot exploit loudness cues from effects like compression or distortion.
For efficiency as in~\cite{steinmetz_st-ito_2024}, we precompute mid/side log-mel spectrograms and AFx-Rep latents.
All PEACE models and classifiers zero-pad these log-mels;
pretrained AFx-Rep keeps its reflection padding.

\subsection{Experimental Setup}
\label{sec:experimental_setup}

We first conduct a hyperparameter search over BoxGraph design choices---Faust parameter masking rate, FFN layers, aggregation function (sum, max, attention, PNA~\cite{corso_principal_2020}), pooling, and more---using precomputed AFx-Rep latents as a fast proxy for end-to-end training.
We then lock the best BoxGraph configuration and other hyperparameters (e.g., learning rate) for the remaining runs.

\textbf{Audio encoder.}
We adopt the AFx-Rep~\cite{steinmetz_st-ito_2024} architecture without changes.
A 77.6M-parameter CNN14 backbone~\cite{kong_panns_2020} processes mid and side channels with shared weights, followed by channel-specific MLP heads.
The two 512-dimensional $L_2$-normalized outputs are concatenated into $y_A \in \mathbb{R}^{1024}$.
We evaluate three initializations (frozen ST-ITO weights, fine-tuned, and scratch).

\textbf{Code encoders.}
The 15.5M-parameter BoxGraph (\S\ref{sec:code_encoder}) operates on the BDA graph.
Alternatively, the 35M-parameter encoder of the smallest T5.1.1~\cite{raffel_exploring_2023,roberts_scaling_2023} takes Faust code with its default tokenizer.
We use general-purpose T5 rather than a code-pretrained encoder such as CodeT5+~\cite{wang_codet5_2023} or GraphCodeBERT~\cite{guo_graphcodebert_2021}, since neither is pretrained on Faust.
We evaluate T5 fine-tuned and from scratch (frozen T5 performed poorly).
For T5, mean-pooling tokens (better than a \texttt{[CLS]} token), rounding floats to two decimals, and keeping parameter comments performed best.
As in T5, we apply 10\% dropout to the FFN, skip connection, attention weights, and module inputs and outputs.
However, we do not use T5-style sentinel tokens for parameter masking.

\textbf{Training.}
Each siamese arm follows SLAP~\cite{guinot_slap_2025} for its projector and predictor: single-hidden-layer MLPs (768-wide for projector; 4096-wide for predictor) with ReLU, and dropout (10\% for projector, 20\% for predictor), with shared dimension 768 for $z$ and $q$.
Including these heads, AFx-Rep-BoxGraph has 108M parameters.
We train for 50,000 steps at effective batch size 256 using AdamW~\cite{loshchilov_decoupled_2019}, weight decay $6 \times 10^{-4}$, 2000 linear warmup steps, peak learning rate $2 \times 10^{-4}$, and cosine decay to $10^{-4}$, selecting models by validation mean reciprocal rank.
On four NVIDIA L40 GPUs, T5 with AFx-Rep from scratch takes about 11.5 hours and BoxGraph 8.5 hours.

\subsection{Evaluation Metrics}
\label{sec:metrics}
We evaluate cross-modal retrieval using Recall@$K$ ($K \in \{1, 10\}$) and median and mean normalized ranks.
Following SLAP~\cite{guinot_slap_2025}, we score retrieval using the predictor outputs $q$ rather than the projector outputs $z$.
To characterize the embedding space beyond retrieval accuracy, we report three measures.
\textit{Gap distance} is the Euclidean distance between the centroids of $q_A$ and $q_C$, where smaller means a tighter shared region.
\textit{Modality separability}~\cite{liang_mind_2022, fahim_its_2024} is the accuracy of a linear probe predicting modality (audio vs.\ code) from $q$.
\textit{Source separability} measures content invariance via one-vs-rest linear probes on $q_A$, one per source type (drums, guitar, singing, speech), reported as mean balanced accuracy across types so that chance remains 50\% despite the 1:3 imbalance.
We use group-aware 5-fold CV keyed on the dry clip filepath so probes cannot exploit per-clip shortcuts.
For both measures, lower is better, since the embedding should encode effects, not content or modality.

\subsection{Room Impulse Response Retrieval Benchmark}
\label{sec:rir_benchmark}
To assess the audio encoder, regardless of cross-modal retrieval, we construct an RIR retrieval benchmark.
We use 1,831 stereo room impulse responses from ReverbFX~\cite{richter_reverbfx_2025}, derived from four professional reverb plugins unrelated to our Faust reverbs.
For dry source audio, we use 250 melodic vocal tracks from the jaCappella corpus~\cite{nakamura_jacappella_2023} (50 songs $\times$ 5 vocal types, excluding vocal percussion).
For each RIR, we convolve two randomly selected but different vocal tracks to produce wet audio.
Given one wet track, the task is to identify which one of 1,831 candidates was processed with the same RIR.
For PEACE models, we evaluate latents ($y_A$) and predictions ($q_A$).
Baselines include the pretrained models AFx-Rep~\cite{steinmetz_st-ito_2024}, Fx-Encoder++~\cite{yeh_fx-encoder_2025}, and LAION-CLAP~\cite{wu_large-scale_2023}.

\begin{table*}[t]
\centering
\small
\resizebox{\textwidth}{!}{
\begin{tabular}{llrrrrrrrrrrr}
\toprule
Audio Encoder & Code Encoder & \multicolumn{4}{c}{Audio $\rightarrow$ Code} & \multicolumn{4}{c}{Code $\rightarrow$ Audio} & Gap ($\downarrow$) & \multicolumn{2}{c}{Separability ($\downarrow$)} \\
\cmidrule{3-6} \cmidrule{7-10} \cmidrule{11-11} \cmidrule{12-13}
& & \multicolumn{2}{c}{Recall ($\uparrow$)} & \multicolumn{2}{c}{Norm. rank ($\downarrow$)} & \multicolumn{2}{c}{Recall ($\uparrow$)} & \multicolumn{2}{c}{Norm. rank ($\downarrow$)} & & & \\
\cmidrule{3-4} \cmidrule{5-6} \cmidrule{7-8} \cmidrule{9-10}
& & @1 & @10 & Med. & Mean & @1 & @10 & Med. & Mean & Dist. & Mod. & Source \\
\midrule
AFx-Rep (scratch) & T5 (scratch) & 41.2 & 71.9 & 0.02 & 1.5 & 41.3 & 71.4 & 0.02 & 1.2 & 0.07 & \textbf{72.2} & \textbf{56.2} \\
AFx-Rep (scratch) & T5 (fine-tune) & \textbf{50.6} & \textbf{78.1} & \textbf{0.00} & 1.1 & 48.4 & 76.3 & 0.02 & 1.0 & 0.07 & 74.0 & 57.0 \\
AFx-Rep (scratch) & BoxGraph & 48.5 & 77.7 & 0.02 & \textbf{1.0} & \textbf{50.8} & \textbf{77.5} & \textbf{0.00} & \textbf{0.8} & \textbf{0.06} & 80.4 & 58.5 \\
\midrule
AFx-Rep (fine-tune) & T5 (scratch) & 46.9 & 74.9 & \textbf{0.02} & 2.5 & 46.9 & 76.1 & \textbf{0.02} & 1.5 & 0.11 & 81.6 & \textbf{54.4} \\
AFx-Rep (fine-tune) & T5 (fine-tune) & 45.0 & 74.5 & \textbf{0.02} & 2.0 & 44.3 & 73.2 & \textbf{0.02} & 1.2 & 0.11 & \textbf{78.1} & 55.9 \\
AFx-Rep (fine-tune) & BoxGraph & \textbf{47.6} & \textbf{77.5} & \textbf{0.02} & \textbf{1.3} & \textbf{48.0} & \textbf{76.3} & \textbf{0.02} & \textbf{1.0} & \textbf{0.09} & 80.3 & 56.6 \\
\bottomrule
\end{tabular}
}
\caption{
Cross-modal retrieval (for metrics, see \S\ref{sec:metrics}).
\textit{Dist.}: Euclidean distance between the $q_A$ and $q_C$ centroids.
\textit{Mod.} and \textit{Source} are linear-probe accuracies (50\% = linearly indistinguishable, 100\% = perfectly separable).
\textbf{Bold} indicates best per column per section. Recall, rank, and separability in \%.
}
\label{tab:encoder_results}
\end{table*}

\subsection{Chain Retrieval}
\label{chain_retrieval}
We evaluate whether PEACE can identify which length-2 effect chain was applied to audio, when all effect parameters are masked.
From 21 Faust effects, we enumerate all 420 ordered pairs of effects, excluding an effect used twice sequentially.
For BoxGraph, we mask all parameter-value embeddings.
For T5, we zero the attention mask at every token inside a parameter literal.
Audio queries are rendered with held out presets on test audio, and retrieval uses cosine similarity of $q_A$ and $q_C$.
We use 1{,}680 ($420 \times 4$) audio queries to span the four source types (\S\ref{sec:dataset}).

As a supervised baseline, we also train a cross-entropy classifier on the identical audio arm, projecting $q_A$ to 441-way logits (effects length 1--2).
It is trained as a 441-way classifier on all 131{,}622 chains, and we score only the 420 pair logits.
We report it at all three AFx-Rep initializations.
A fourth variant instead warm-starts its audio arm from the AFx-Rep (scratch) BoxGraph model.
Prior chain classification methods~\cite{guo_automatic_2023, wada_hyperbolic_2025, rossi_automatic_2025, okita_audio_2026} produce labels rather than embeddings, so none work automatically on our Faust topology gallery.

\section{Results}

We evaluate PEACE on cross-modal retrieval, per-effect retrieval, chain-length generalization, reverb matching, and chain retrieval.
Overall, PEACE demonstrates both cross-modal and intramodal understanding, consistent with its SLAP objective.
Performance varies across effect types, revealing limitations of the current audio representation and suggesting directions for future improvement.

\tabref{tab:encoder_results} reports cross-modal retrieval for effect chains of length 1--3.
With AFx-Rep trained from scratch, fine-tuned T5 marginally leads on Audio$\rightarrow$Code retrieval, while BoxGraph marginally leads on Code$\rightarrow$Audio.
For slightly fairer comparison, requiring T5 to be trained from scratch makes it underperform BoxGraph.
Still, the results suggest T5 trained end-to-end can partially overcome the tokenization pitfalls evident in prior work~\cite{kouteili_embedding_2025}, at least when floats are rounded to two decimals.

The modality gap metrics are consistent with SLAP's Figure~3~\cite{guinot_slap_2025}.
With AFx-Rep from scratch, pairings with T5 produce less separable modality embeddings than BoxGraph.
All end-to-end trained models may have similar source separability due to the shared audio encoder architecture.
Source separability in the mid-50s is close to the 50\% chance level, indicating the embeddings are largely invariant to source content, a desirable property.

\subsection{Per-Effect Retrieval Results}
\label{sec:per_effect_results}
\tabref{tab:per_effect_retrieval} reports bidirectional retrieval \emph{within} each effect type (length-1 chains).
Overall, T5 marginally leads on both R@1 and R@10.
On R@10, BoxGraph wins 11 of 21 effects, including dynamics and non-linear effects (\textit{Compressor}, \textit{Distortion}, \textit{Gate}, \textit{Limiter}), while T5 wins most time-based effects such as \textit{Chorus}, \textit{Delay}, \textit{Flanger}, \textit{Phaser}, \textit{ReverseEcho}, and three of five reverbs.

\begin{table}[t!]
\centering
\small
\begin{tabular}{lrr|rr}
\toprule
 & \multicolumn{2}{c|}{R@1 ($\uparrow$)} & \multicolumn{2}{c}{R@10 ($\uparrow$)} \\
\cmidrule(lr){2-3} \cmidrule(lr){4-5}
Effect & Box & T5 & Box & T5 \\
\midrule
AutoPan & \textbf{20.6} & 19.5 & \textbf{62.8} & 60.7 \\
Chorus & 27.7 & \textbf{34.5} & 74.4 & \textbf{78.3} \\
Compressor & 2.5 & \textbf{2.6} & \textbf{16.8} & 14.6 \\
Delay & 31.2 & \textbf{35.1} & 71.4 & \textbf{74.2} \\
Distortion & \textbf{4.6} & 3.7 & \textbf{21.0} & 20.8 \\
EQ & \textbf{5.7} & 4.8 & \textbf{23.1} & 19.2 \\
Filter & \textbf{6.2} & 4.6 & \textbf{22.1} & 20.1 \\
Flanger & 29.5 & \textbf{31.8} & 74.1 & \textbf{77.1} \\
Gate & \textbf{0.3} & 0.1 & \textbf{2.7} & 2.3 \\
KlonCentaur & \textbf{3.0} & 2.1 & \textbf{18.7} & 16.0 \\
Limiter & \textbf{0.5} & 0.2 & \textbf{2.8} & 2.4 \\
Panner & 1.1 & \textbf{2.8} & 7.9 & \textbf{8.6} \\
Phaser & \textbf{28.8} & 25.1 & 68.3 & \textbf{68.9} \\
PitchShift & \textbf{11.5} & 9.8 & \textbf{48.3} & 44.1 \\
ReverbDattorro & 29.5 & \textbf{34.6} & 70.8 & \textbf{76.7} \\
ReverbFreeverb & \textbf{43.5} & 42.4 & \textbf{84.5} & 84.1 \\
ReverbSpring & 34.5 & \textbf{35.7} & \textbf{74.8} & 74.4 \\
ReverbVital & 18.0 & \textbf{26.3} & 52.8 & \textbf{64.2} \\
ReverbZita & 45.1 & \textbf{54.8} & 83.4 & \textbf{87.7} \\
ReverseEcho & 7.9 & \textbf{9.2} & 49.2 & \textbf{54.1} \\
StereoWidth & \textbf{0.2} & \textbf{0.2} & 0.9 & \textbf{1.0} \\
\midrule
\textbf{Overall} & 16.8 & \textbf{18.1} & 44.3 & \textbf{45.2} \\
\bottomrule
\end{tabular}

\caption{
    Per-effect average retrieval in a gallery of size 1024, in \%. Chance is $\sim$0.1\% R@1 and $\sim$1.0\% R@10.
}
\label{tab:per_effect_retrieval}
\end{table}

Putting code encoders aside, the variance across effects offers a roadmap for improving the audio encoder and training data.
Time-based effects such as \textit{Reverb} and \textit{Delay} have the highest retrieval performance.
This may reflect AFx-Rep's CNN input of log-mel spectrograms.
Non-linear and dynamics-based effects such as \textit{Compressor} and \textit{Distortion} would likely benefit from time-domain encoders.
\textit{Panner} needs a frontend that tells left from right: AFx-Rep's mid/side spectrograms are identical whether a source is panned left or equally far right.

For certain parameters and inputs, some effects place a hard ceiling on retrieval.
\textit{Limiter} and \textit{Gate} may not change audio for quiet and loud inputs, respectively.
Similarly, \textit{StereoWidth} doesn't change the width of mono inputs.

\subsection{Chain Length Generalization}
\label{sec:chain_length}

\begin{figure}[t!]
\centering
\includegraphics[alt={Line plot of R@1 and R@10 retrieval recall vs.\ chain length N from one to eight, for BoxGraph and T5 code encoders.
T5 and BoxGraph are essentially tied inside the shaded training range (N up to 3), but T5 generalizes better beyond it: BoxGraph R@10 falls from the mid-90s at N=3 to the low 60s at N=8, while T5 R@10 stays above 85 for every N beyond 3.
},width=\columnwidth]{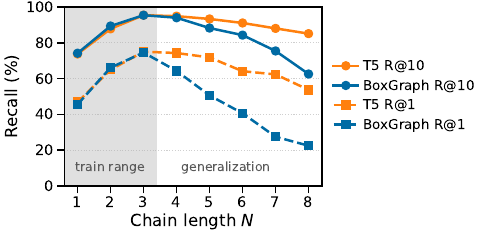}
\caption{
    Retrieval vs.\ chain length $N$ (gallery size 1024).
    Shaded region marks chain lengths seen during training.
}
\label{fig:per_num_effects_retrieval}
\end{figure}

\figref{fig:per_num_effects_retrieval} shows bidirectional retrieval as chain length grows, extending to $N \in \{1,\ldots,8\}$ the axis of ST-ITO's Figure~4~\cite{steinmetz_st-ito_2024}, which breaks down audio-to-audio style retrieval for $N \in \{1,\ldots,5\}$.

At $N=1$, recall is higher than in \tabref{tab:per_effect_retrieval}, for the same gallery size.
This makes sense because differences in effect types should be more salient than changing parameters within an effect.
The encoders track each other closely for $N\!\leq\!3$, peaking at $N\!=\!3$, the end of the training distribution.
We speculate the task shifts from identifying parameters towards identifying effects.
Beyond the training distribution, T5 degrades gradually and BoxGraph much faster.
Training on longer chains might remedy both.

\subsection{Reverb Retrieval Results}
\label{sec:rir_results}

\tabref{tab:rir_benchmark} reports the RIR retrieval benchmark (\S\ref{sec:rir_benchmark}), where audio comes from professional plugins unrelated to Faust and unseen in training.
End-to-end PEACE training achieves intramodal understanding of audio.
PEACE scratch BoxGraph $q_A$ reaches 53.5\% R@10, far exceeding pretrained AFx-Rep (28.3\%) and the CLAP-based Fx-Encoder++ (11.6\%) and LAION-CLAP (10.7\%).
This gap echoes prior research showing that joint language-audio embeddings weakly encode effect-relevant timbre~\cite{deng_joint_2025}.

For PEACE, the predictor $q_A$ always outperforms the latent $y_A$.
With frozen AFx-Rep latents and BoxGraph, R@10 rises from 28.7\% to 38.6\%.
This suggests cross-modal alignment via SLAP can inexpensively adapt representations from frozen effects encoders for improved understanding of reverb and possibly other effects.
Another motivation for using $q_A$ instead of $y_A$ for retrieval or classification comes from CLIP's Figure~5 in which zero-shot CLIP embeddings outperform few-shot linear probes on pre-embedding CLIP latents~\cite{radford_learning_2021}.

\begin{table}[t!]
\centering
\small
\begin{tabular}{llrrrr}
\toprule
& & \multicolumn{2}{c}{R@1 ($\uparrow$)} & \multicolumn{2}{c}{R@10 ($\uparrow$)} \\
\cmidrule(lr){3-4} \cmidrule(lr){5-6}
Model & Code & $y_A$ & $q_A$ & $y_A$ & $q_A$ \\
\midrule
PEACE scratch & BoxGraph & \textbf{19.4} & \textbf{23.1} & \textbf{50.6} & \textbf{53.5} \\
PEACE scratch & T5 & 18.4 & 19.1 & 46.5 & 48.3 \\
PEACE frozen & BoxGraph & 8.1 & 13.9 & 28.7 & 38.6 \\
PEACE frozen & T5 & 8.1 & 13.4 & 28.7 & 36.6 \\
\midrule
AFx-Rep & -- & 8.4 & -- & 28.3 & -- \\
Fx-Encoder++ & -- & 2.3 & 2.3 & 11.6 & 10.9 \\
LAION-CLAP & -- & 1.8 & 1.8 & 10.7 & 10.2 \\
\bottomrule
\end{tabular}

\caption{
    RIR retrieval (gallery 1,831), in \%.
    $y_A$ = encoder latent (HTS-AT for LAION-CLAP);
    $q_A$ = predictor output, or the projector output $z_A$ for Fx-Encoder++ and LAION-CLAP, which lack a predictor.
}
\label{tab:rir_benchmark}
\end{table}

\subsection{Chain Retrieval Results}
\label{sec:chain_identification}

On the chain retrieval benchmark (\S\ref{chain_retrieval}), PEACE succeeds without a supervised classification objective (\tabref{tab:chain_retrieval}).
AFx-Rep-BoxGraph reaches 17.4\% R@1 and 61.4\% R@10, beating T5's 12.4\% R@1 and 43.2\% R@10.
Counting a retrieval as correct when the top-$K$ contains any chain with the same effect set raises BoxGraph to 27.2\% and 67.4\% (T5, 20.7\% and 49.1\%), partly because LTI effects commute without changing the output~\cite{smith_introduction_2007}.

Turning to the supervised rows, chain labels help only when the audio backbone learns.
With frozen AFx-Rep, the classifier ties BoxGraph (12.7\% against 12.6\% R@1), and letting the encoder train from scratch raises it to 21.4\%.
Applying the same supervised training signal to an audio backbone warm-started from a strong PEACE checkpoint yields a nearby 21.1\% R@1 and converges faster.

Although the best retrieval row lags behind the best supervised row, retrieval could scale to much longer chains (\figref{fig:per_num_effects_retrieval}).
Supervised methods\cite{guo_automatic_2023, wada_hyperbolic_2025, rossi_automatic_2025, okita_audio_2026} remain limited to a fixed set of labels seen in training.

\section{Conclusion}

This paper presents PEACE, the first joint embedding of DSP effects code and effected audio.
Two distinct code encoders, fine-tuned T5 and BoxGraph, are competitive on retrieval of mixed-length chains.
T5 holds a minor lead on average within-effect retrieval, but BoxGraph wins about half of the effect types.
T5 generalizes better to long chain lengths, while BoxGraph wins order-aware chain retrieval from audio.
All PEACE models do well on intramodal reverb retrieval, particularly BoxGraph.
Interestingly, cross-modal SLAP training on a pretrained frozen AFx-Rep improves performance on this intramodal task.

\textbf{Limitations.}
As in prior work~\cite{steinmetz_st-ito_2024, yeh_fx-encoder_2025}, three aspects of our training pipeline widen the sim-to-real gap.
Parameter values are randomly sampled rather than human-curated;
presets are randomly paired with source audio;
and the source audio itself is not always fully dry.
The variance in within-effect retrieval warrants improvements in data quality and the audio encoder.
Our out-of-domain evaluation covers only reverb rather than a broader range of plugin effects, and we lack a user study.

\begin{table}[t]
\centering
\small
\setlength{\tabcolsep}{5pt}
\begin{tabular}{lrr|rr}
\toprule
 & \multicolumn{2}{c|}{Exact ($\uparrow$)} & \multicolumn{2}{c}{Order-tolerant ($\uparrow$)} \\
Method & R@1 & R@10 & R@1 & R@10 \\
\midrule
\multicolumn{5}{l}{\textit{Retrieval (no chain supervision)}} \\
AFx-Rep frozen + Box & 12.6 & 48.7 & 19.5 & 56.8 \\
AFx-Rep scratch + Box & 17.4 & 61.4 & 27.2 & 67.4 \\
AFx-Rep scratch + T5 & 12.4 & 43.2 & 20.7 & 49.1 \\
\midrule
\multicolumn{5}{l}{\textit{Supervised classifier (cross-entropy)}} \\
AFx-Rep frozen & 12.7 & 50.7 & 21.0 & 61.1 \\
AFx-Rep scratch & \textbf{21.4} & \textbf{65.9} & \textbf{32.6} & \textbf{73.6} \\
AFx-Rep fine-tune & 20.2 & 65.2 & 31.4 & 72.3 \\
PEACE Box init & 21.1 & 62.9 & 31.6 & 70.0 \\
\bottomrule
\end{tabular}

\caption{
Chain topology retrieval in \% from 420 ordered pairs of distinct effects with all parameter values masked.
Chance is 0.24\% R@1 and 2.4\% R@10.
Supervised rows use only an audio encoder and average three seeds, which differ by as much as 4.9 points in R@1.
Order-tolerant counts a retrieval as correct if the top-$K$ contains any chain with the same set of effects regardless of order.
}
\label{tab:chain_retrieval}
\end{table}

\textbf{Future work.}
BoxGraph can extend to new terminal DSP nodes for modular synthesis (e.g., envelopes and LFOs) and to the remaining composition operators (\texttt{split}, \texttt{merge}, and \texttt{recurse}) for richer DSP graphs beyond effect chains.
However, the Bitter Lesson~\cite{sutton_bitter_2019} would favor general-purpose code encoders such as multimodal LLM architectures.

PEACE embeddings could be the target space for style transfer~\cite{steinmetz_st-ito_2024, yu_improving_2025} or condition multi-track mixing~\cite{moliner_automatic_2026}.
In the pipeline of~\cite{okita_audio_2026}, our chain retrieval method (\S\ref{chain_retrieval}) could replace the DNN effect classifier: given a target wet signal, retrieve the top-K chain topologies, then hand these candidates to CMA-ES parameter search.
In the same pipeline, PEACE embeddings could condition the effect removal network.
Finally, PEACE's code encoder could speed up parameter search by reducing the burden of iteratively rendering audio.

\section{Acknowledgments}

This work was supported in part by the National Science Foundation under award number IIS-2523649.

\bibliography{references}

\end{document}